\documentstyle{l-aa}
\begin{document}

\title{Optical observations of BL Lacertae from 1997 to 1999}

\author{J.H. Fan$^{1,2,3}$ \and B.C. Qian$^{4}$ \and J. Tao$^{4}$}
 \institute{ Center for Astrophysics, Guangzhou University,
 Guangzhou 510400, China, e-mail: jhfan@guangztc.edu.cn
\and Chinese Academy of Science-Peking University Joint Beijing
Astrophysical Center(CAS-PKU.BAC), Beijing, China
\and Department of Physics, Yunnan University,
Kunming, China
\and Shanghai Astronomical Observatory, Chinese Academy of Sciences,
Shanghai, China }
\offprints{J.H. Fan}
\maketitle
\begin{abstract}

 We present the optical (V, R, and I) photometry for BL Lacertae,
 which was observed from 1997 through 1999, with the 1.56-m telescope 
 at the Shanghai astronomical observatory (SHAO).  After the
1997 outburst, it dimmed
to a low state and then brightened again.  During the period 
JD 2450701 to JD 2450701.5, variations of 0.40mag, 0.27mag, and
0.21mag
over a time scale of 100 minutes were found for V, R, and I bands,
suggesting that the variations were decreasing with wavelength.
The  correlation between V, R, and I is also analyzed using the
DCF (Discrete Correlation Function) method. This shows that
the variability in the V, R, and I bands are correlated with
no time delay longer than 0.2 day.
\end{abstract}


\section{Introduction}

The nature of active galactic nuclei (AGNs) is still an open problem.
Photometric observations of AGNs are important for constructing their
light curves and to study their variation behavior on different
 time scales.  Blazars are an extreme subclass of  AGNs and often
 show large and violent variations. The variability on very
 short time scales from minutes to hours is  a common property
 of blazars, 
 and is observed in many objects
 (
 Rieke et al. 1976;
 Smith et al. 1987;
 Sillanpaa et al. 1991;
 Carini et al. 1992;
 Romero et al. 1994, 1995a,b, 1997, 2000;
 Heidt \& Wagner 1995;
 Miller \& Noble 1996;
 Villata et al. 1997, 1999;
 Terasranta et al. 1998;
 Takalo 1994;
 Bai et al. 1999;
 Kraus et al. 1999;
 Raiteri et al. 1999).

 BL Lacertae (PKS 2200+420), the archetype of its class, lies in a
 giant elliptical galaxy at a redshift of $\sim 0.07$ (
 Miller et al. 1978). It is one of the best-studied objects.
 Superluminal components have been observed from the source(
 Mutel \& Phillips 1987;
 Vermeulen \& Cohen 1994;
 Fan et al. 1996;
 Xie et al., 1992, 1994;
 Webb et al. 1988; 1998;
 Catanese et al. 1997;
 Qin \& Xie 1997;
 Fan et al. 1998a,b;
 Fan \& Lin 1999, 2000
 and reference therein). Its spectrum is  usually featureless,
 but weak emission lines are indeed identified when the source
 is in the fainter state (Corbett et al. 1996).
 The optical variability is extremely irregular over periods of hundreds
 of  days of continuous observation but a possible $\sim$ 14-year period
 was found in the B light curve (see Fan et al. 1998a).
 In addition, some rapid variability over short time scales has been
 reported: for example, a variation of 1.5 magnitude over a time scale
 of 20 hours (Weistrop, 1973), daily variation as great as 0.3 magnitude
 (Carswell et al. 1974),  
 and variation of 0.1 magnitude over 30 minutes in the V band
 (Corbett et al. 1996).
 During the 1997 outburst period, it was widely observed in optical bands
 by many groups (
 Nesci et al. 1998;
 Bai et al. 1999;
 Kovael et al. 1999;
 Massaro et al. 1999;
 Matsumoto et al. 1999;
 Miller et al. 1999;
 Nikolashvili et al. 1999a,b;
 Sobrito et al. 1999;
 Tosti et al. 1999a,b;
 Qian et al. 2000;
 Ghosh et al. 2000a,b
 and reference therein).

 BL Lacertae is also one of the sources in our observing program.
 To investigate their short time-scale variability, 
 we have made optical
 observations of blazars with the 1-m telescope at Yunnan
 Astronomical Observatory (YAO),
 the 1.56-m telescope at Shanghai Observatory (SHAO), and
 the 2.16-m  telescope at Beijing Observatory (BAO)
 (Fan et al. 1997, 1998a; Qian et al. 2000; Xie et al. 1992, 1994).

 In this paper, we present the BL Lacertae optical photometry
  observed with
  the 1.56-m telescope at SHAO during the period  1997 to 1999.
  In section 2, we present
  the data and analysis; in section 3, a brief conclusion.

\section{Observations and analysis}
\subsection{Data reductoin}

 BL Lac is one of the regularly monitored objects at SHAO
 using the $f/10$ Cassegrain focus of the 1.56m telescope with a
 liquid nitrogen cooled Photometric 200 series CCD camera having
 $1024\times1024$ pixels.
 The filters are standard Johnson and Cousins BVRI filters. The
 field of view is 4'17", with 1 pixel=0.25".  The 
 exposure times are typically set according to the sky conditions.
 The seeing at the Sheshan Station of SHAO usually varies from
 1.2" to 1.5".

 The obtained frames were processed with the
 photometric task APPHOT of the IRAF software package after bias,
 dark and flat-field corrections. The bias frames were taken at the
 beginning and the end of the night observation. In addition, some were 
 taken in the middle of the observations. Sky flat-field images
 were taken at dusk and dawn where possible. Otherwise,
 a dome flat was used. 
 The dark and the flat-field corrections are 0.01 to 0.02 mag, 
 mainly contributed  by the flat-field. 

 We used stars B and C (Smith et al., 1985)
 as photometric comparison  stars. Their colors are
 $(V-R) = 0.85 \pm 0.06$,  $(R-I) = 0.84 \pm 0.08$ for B
 and $(V-R) = 0.50 \pm 0.04$,  $(R-I) = 0.46 \pm 0.05$ for C
 while their magnitudes are 
 $V = 12.78 \pm 0.04$, $R = 11.93 \pm 0.05$, $I = 11.09 \pm 0.06$ for B
 and
 $V = 14.19 \pm 0.03$, $R = 13.69 \pm 0.03$, $I = 13.23 \pm 0.04$ for C.
 Based on  the long-term data, the colors of BL Lacertae are 
 $(V-R) = 0.73 \pm 0.19$ and 
 $(R-I) = 0.82 \pm 0.11$ on average, while the magnitudes are in the
 ranges: V = 10.52 to 15.25, R = 11.78 to 14.91, and
 I = 10.74 to 14.19 (Fan et al. 1998a). 
 Because the colors of the comparison stars are similar to 
 those of BL Lacertae while the magnitudes of the comparison stars
 are in the magnitude range of BL Lacertae, the reduction of the 
photometry  is simplified, with no needed corrections for color
or non-linearity.

 We determined  differential magnitudes of BL-B and B-C from 
 the instrumental
 magnitudes of BL Lacertae (BL), the Star B (B), and star C (C). 
 The curves B-C indicate observational uncertainties and  the intrinsic
variability of the stars. 
 The variability of the target object BL Lacertae is 
 investigated by means of the variability parameter, $C$,
 introduced by Romero et al. (1999, see also Cellone et al. 2000). 
 To do so, we determine the 
 scatter of the differential magnitudes  BL-B and B-C,
 $\sigma_{(BL-B)}$ and 
 $\sigma_{(B-C)}$, the variability parameter $C$ is expressed as
 ${\frac{\sigma_{(BL-B)}}{\sigma_{(B-C)}}}$.
 If $C~>~3$, then the target is variable. 

 The rms errors  are calculated from the two stars using 
 the formula:
 $$\sigma~=~\sqrt{{\frac{\Sigma(m_{i}-\overline{m})^{2}}{N-1}}}$$
 where $m_{i}~=~(m_{B}-m_{C})_{i}$ is the differential magnitude 
 of stars B and C
 while $\overline{m}~=~\overline{m_{B}-m_{C}}$ is the  
 differential magnitude averaged over the entire dataset, and $N$
 is the number of the observations on a given night.
 The results are given in Table 1 for filters V, R and I.
 Column (1) is the Julian date,
 column (2) the V magnitude,
 column (3) the uncertainty in V,
 column (4) the R magnitude,
 column (5) the uncertainty in R,
 column (6) the I magnitude, and
 column (7) the uncertainty in I.
 The light curves are shown in Figure 1 for the bands R, V and I.

\subsection{Analysis}
\subsubsection{Variation}

 During the 1997 outburst, rapid variations of $0^{m}.5$ over 90
 minutes by Sobrito et al. (1999) and $0^{m}.6$ over 40 minutes
 by Matsumoto et al. (1999) were found (see also Nesci et al. 1998).
 Our observations show that in the period JD 2450692 to 
 JD 2450702,
 a variation of about 1 mag over a week is found in the three
 bands. In the period  JD 2450701 to JD 2450701.5, variations of
 0.40mag, 0.27mag, and 0.21mag over a time scale of 100 minutes
 were found in the V, R, and I wavebands (see Fig. 2) suggesting that the
variations decrease with increasing wavelength,  consistent with the 
 findings of Nikolashvili et al. (1999a). No similar rapid variation
 was found in our other observations of the source during our monitoring
 period 1997 to 1999, which suggests that this does not happen often. 
 The corresponding variability parameters $C_{V,R,I}$ are greater 
 than 10.0. 

 After the 1997 outburst, it dimmed to a very low state  (Fig. 1),
 with R = 14.76 on JD 2450738, and then brightened again. The brightening
 tendency is consistent with the later observation of a high
 state of R = 12.44 on JD 2451514.32 (Massaro \& Nesci 1999).
 On JD 2451328.5  R = 13.10 was observed by
 Tosti \&  Nucciarelli (1999). On those days, we have no observations,
 but our data (R = 13.24 on JD 2451104 and R = 13.18 on JD 2451379)
 combined with those by  Tosti \&  Nucciarelli (1999), and 
 Massaro \& Nesci (1999) suggest a  brightening tendency.

\subsubsection{Correlated variability}

 BL Lacertae was observed in V, R, and I bands from 1997 through 1999,
 which makes it possible for us to discuss the correlated variability among
 the colors. We analyzed our time-series data to search for time lag
 using the method of Discrete Correlation Function (DCF)
 (Edelson \& Krolik 1988; also see  Fan et al. 1998c, and
 Tornikoski et al. 1994). 

 First, we  calculated the set of
 unbinned correlation coeffieient (UDCF) between data points in the two
data streams $a$ and $b$, i.e.
\begin{equation}
{UDCF_{ij}}={\frac{ (a_{i}- \bar{a}) \times (b_{j}- \bar{b})}{\sqrt{\sigma_{a}^2 \times \sigma_{b}^2}}},
\end{equation}
where $a_{i}$ and $ b_{j}$ are points in the data sets, $\bar{a}$
and $\bar{b}$ are the average values of the data sets, and $\sigma_{a}$
and $\sigma_{b}$ are the corresponding standard deviations. Secondly,
we averaged the points sharing the same time lag by
binning the $UDCF_{ij}$ in suitably sized time-bins in order to get the
$DCF$ for each time lag $\tau$:
\begin{equation}
{DCF(\tau)}=\frac{1}{M}\Sigma \;UDCF_{ij}(\tau),
\end{equation}
where $M$ is the total number of pairs. The standard error for each bin is

\begin{equation}
\sigma (\tau) =\frac{1}{M-1} \{ \Sigma [ UDCF_{ij}-DCF(\tau)]^{2}\}^{0.5}.
\end{equation}
The results for time bins of 0.2 days are shown in Fig. 2 for V vs. R, R
vs. I, and V vs. I respectively. No time delay longer than 0.2 day 
is found between any two bands.

\section{Conclusion}

BL Lacertae is a variable source through the whole magnetic waveband,
and has been observed intensively. 
In this paper, we presented our measurements of V, R, and I bands 
for the period of 1997 to 1999. Short time scale variability over
$\sim$ 100 minutes was found in the three bands, with 
the variation amplitude found to decrease with  wavelength. This
variation property was also noted by other authors.
The optical variations are found to be
correlated with no time delay exceeding 0.2 days.

\section*{Acknowledgements} 
The authors thank Dr. Wills for the comments and suggestions that
improve the paper. JHF thanks Dr. G.E. Romero for his comments.
This work is supported by the National Scientific
Foundation of China (19973001) and the National 973 Project of 
China (NKBRAF G19990754).


{}

\newpage
\begin{figure*}
\vbox to7.2in{\rule{0pt}{7.2in}}
\includegraphics{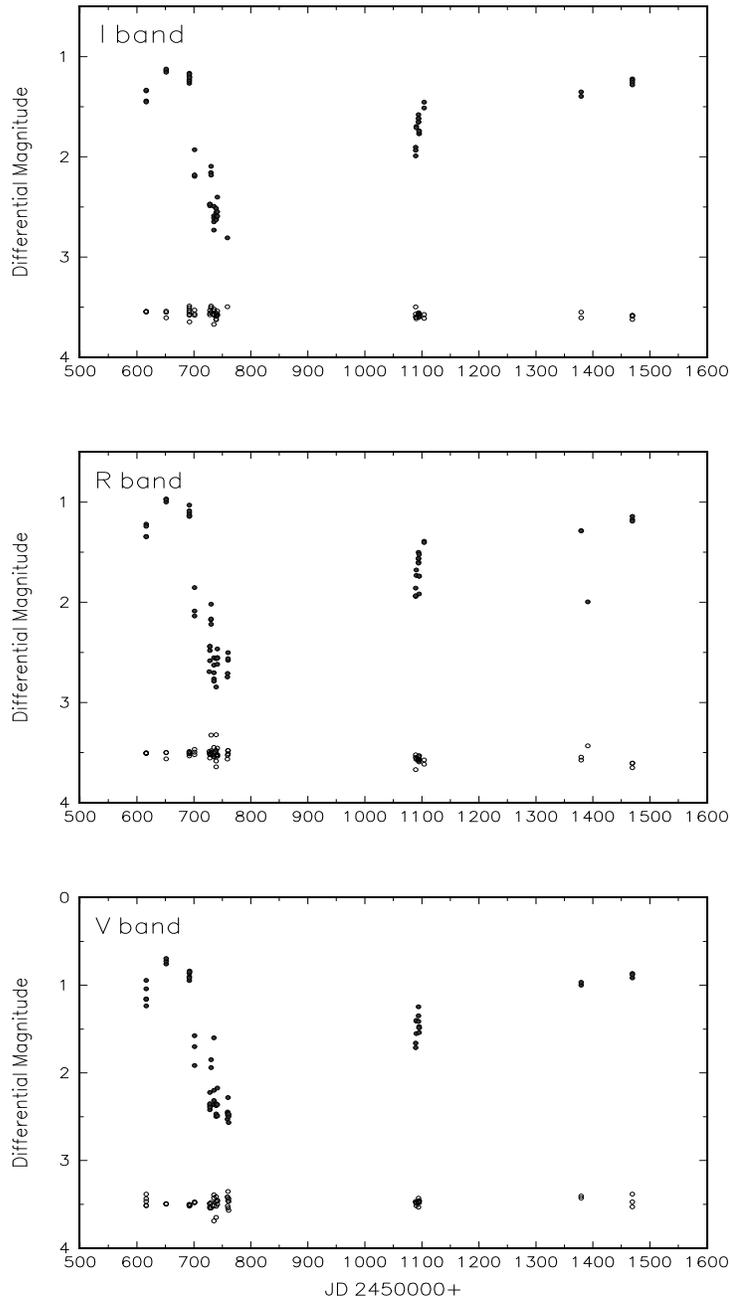}
\caption{ 
 Differential Light Curves of BL-B and B-C in I band (upper 
panel), in R band (middle panel), and in V band (lower panel). The light
curves of B-C are displaced by 3.5 mag. for all the three bands.
}
\end{figure*}

\newpage
\begin{figure*}
\vbox to7.2in{\rule{0pt}{7.2in}}
\includegraphics{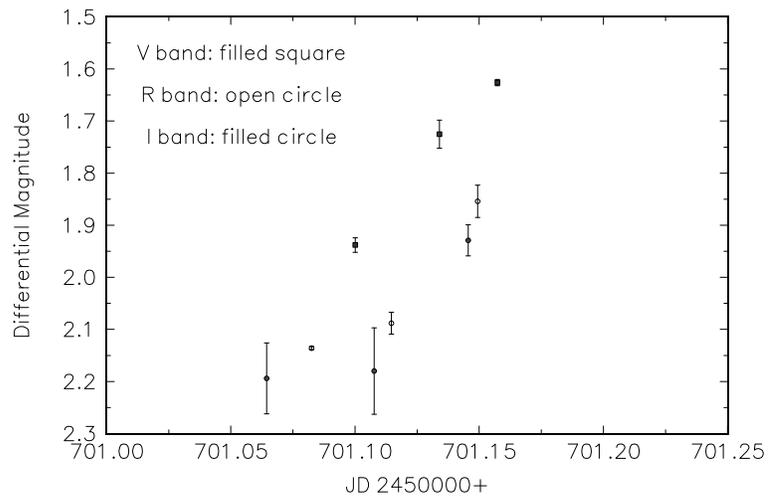}
\caption{ Variation on JD 2450701 in V (filled squares), R
(open circles), and I (filled circles)} bands.
\end{figure*}

\newpage
\begin{figure*}
\vbox to7.2in{\rule{0pt}{7.2in}}
\includegraphics{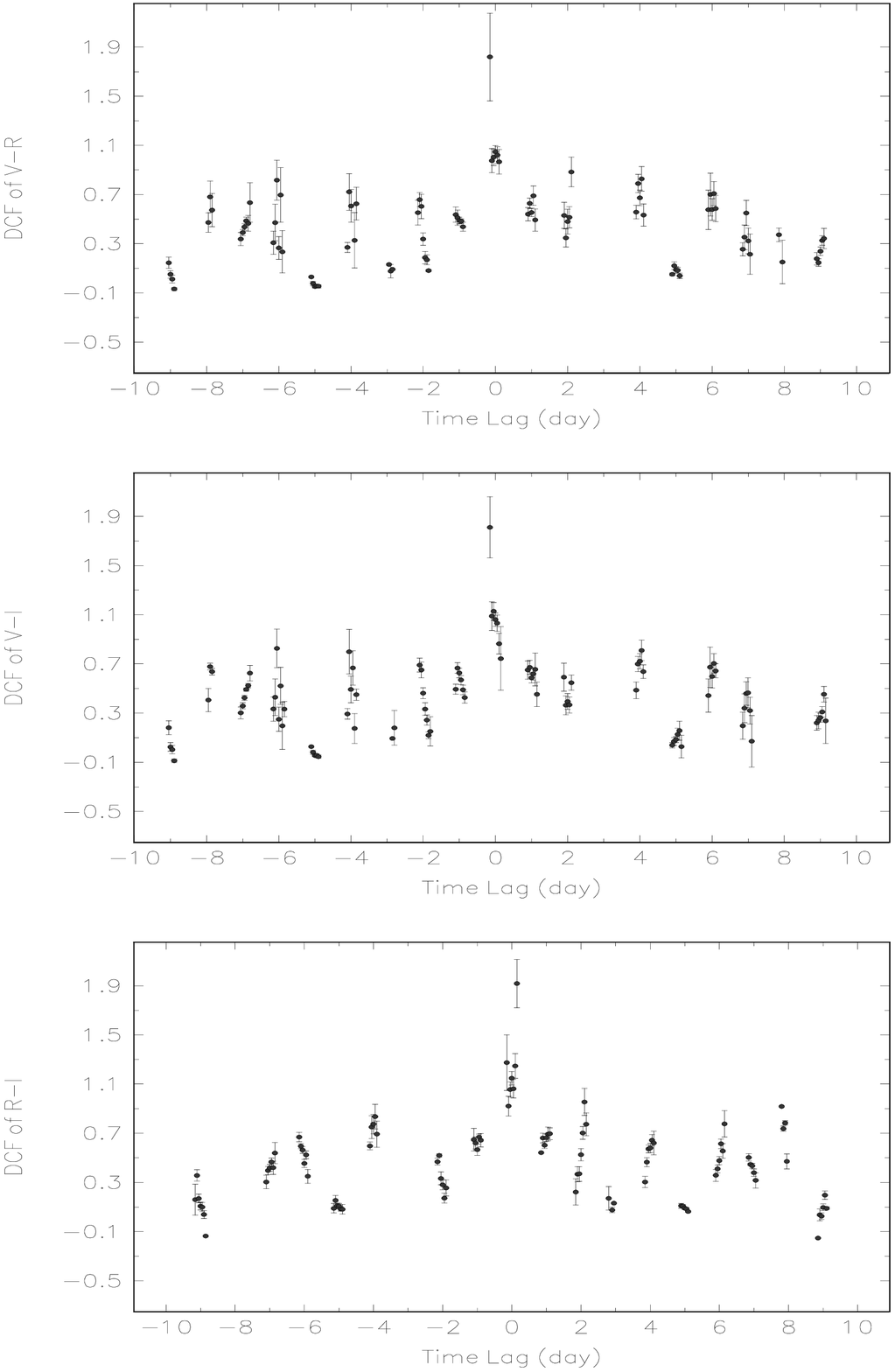}
\caption{Plot of DCF V-R (upper panel), DCF R-I (middle
panel), and DCF V-I (lower panel).}
\end{figure*}

\clearpage
\begin{figure*}
\vbox to7.2in{\rule{0pt}{7.2in}}
\includegraphics{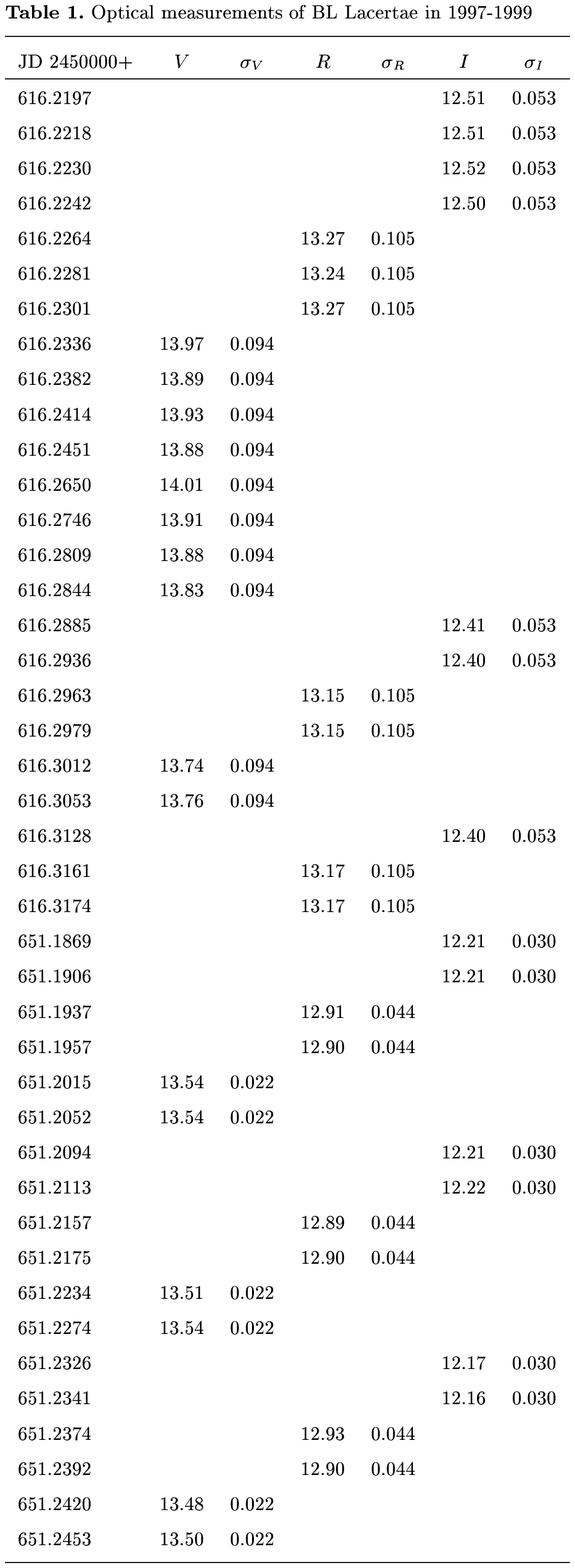}
\end{figure*}

\begin{figure*}
\vbox to7.2in{\rule{0pt}{7.2in}}
\includegraphics{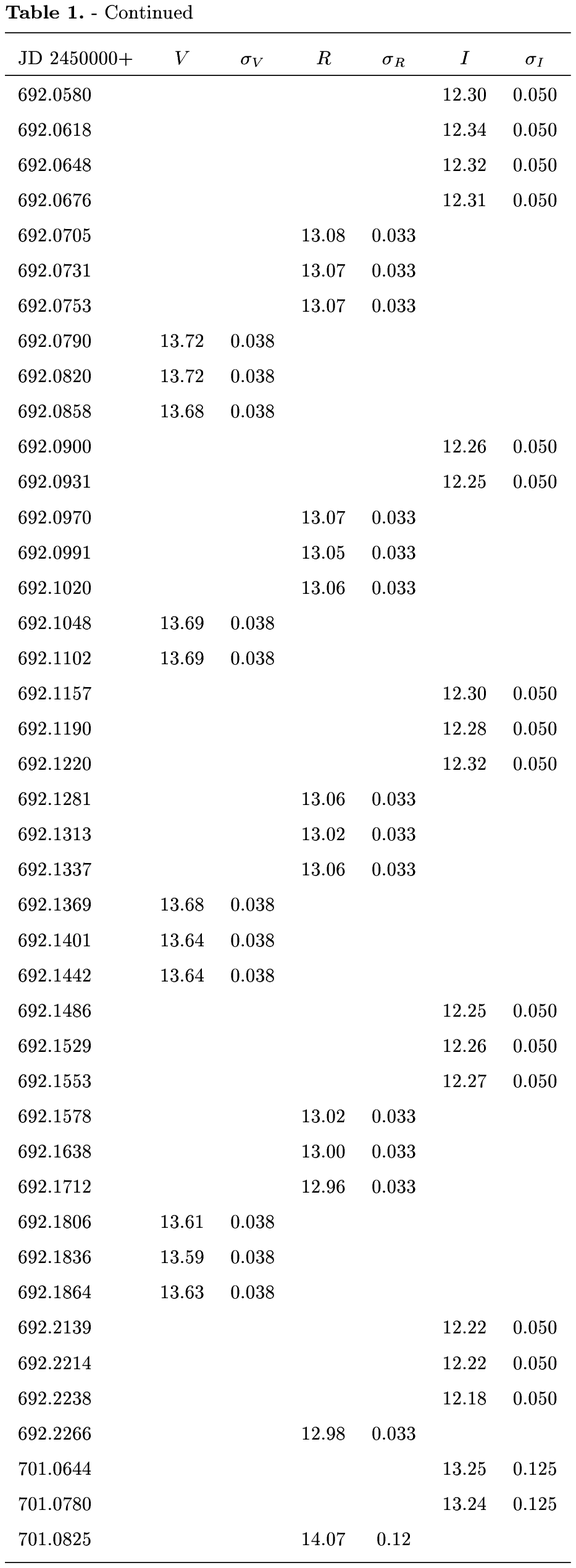}
\end{figure*}
\begin{figure*}
\vbox to7.2in{\rule{0pt}{7.2in}}
\includegraphics{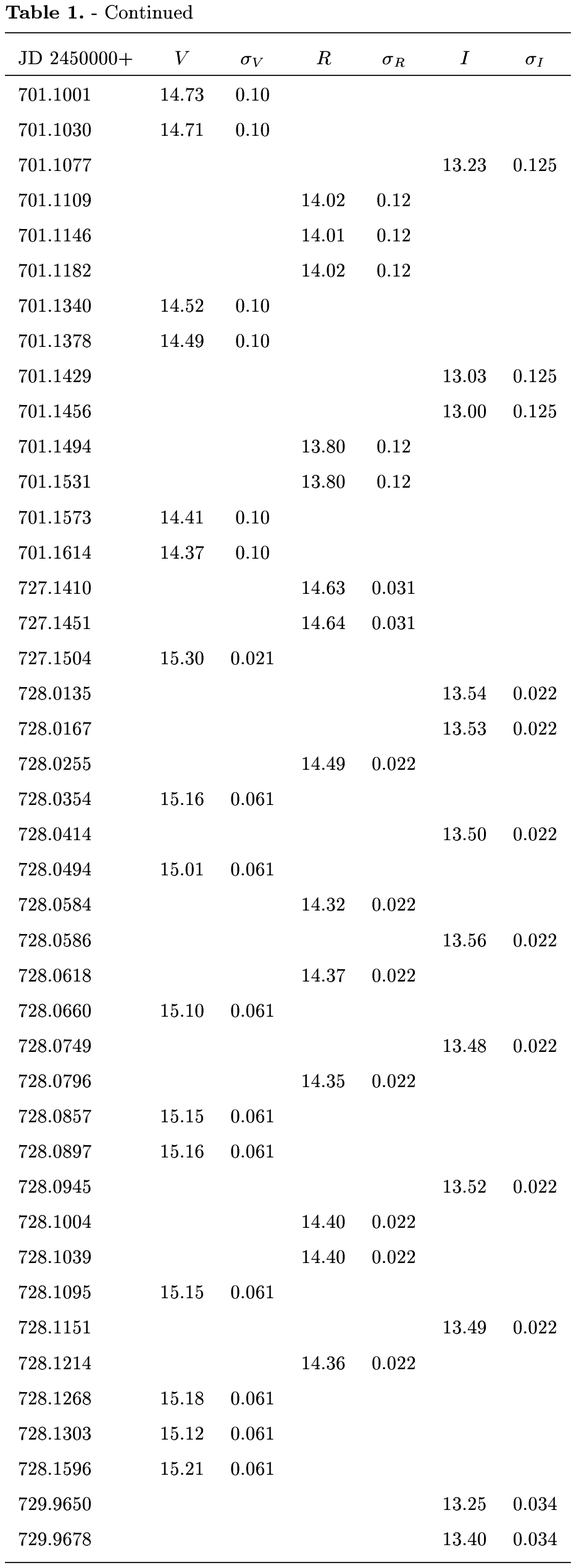}
\end{figure*}

\begin{figure*}
\vbox to7.2in{\rule{0pt}{7.2in}}
\includegraphics{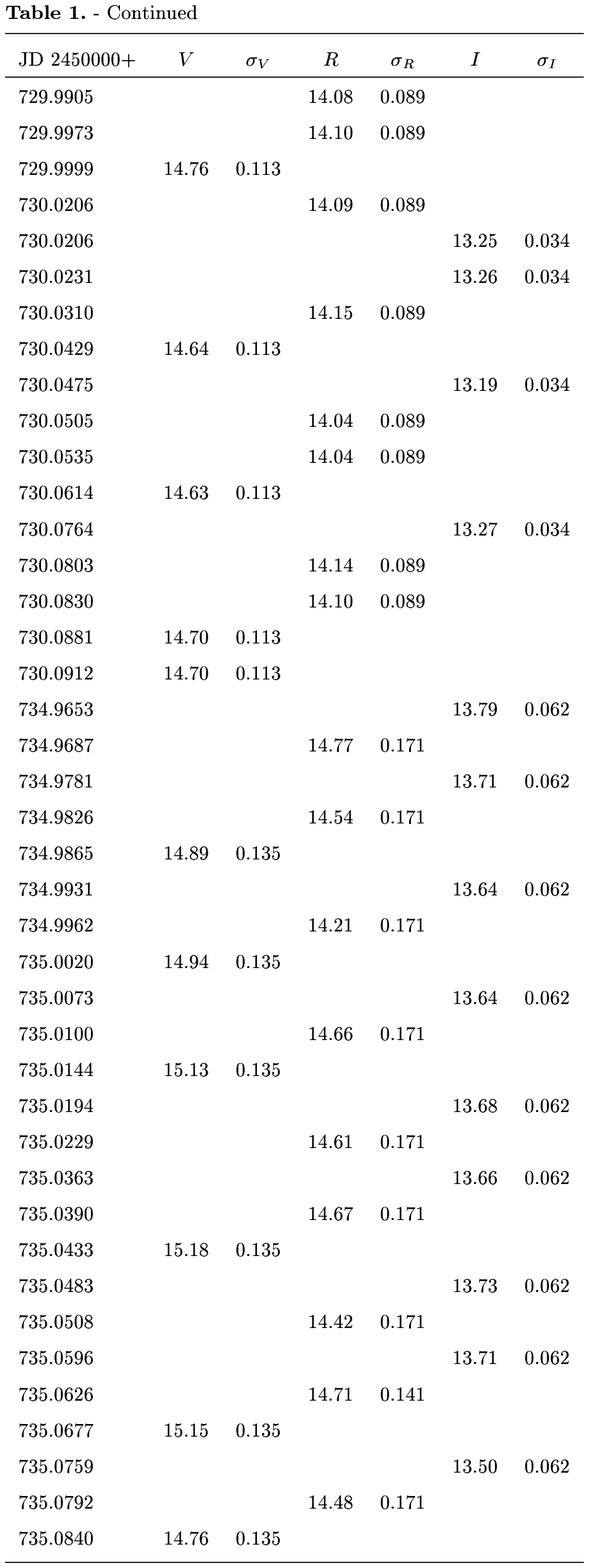}
\end{figure*}
\begin{figure*}
\vbox to7.2in{\rule{0pt}{7.2in}}
\includegraphics{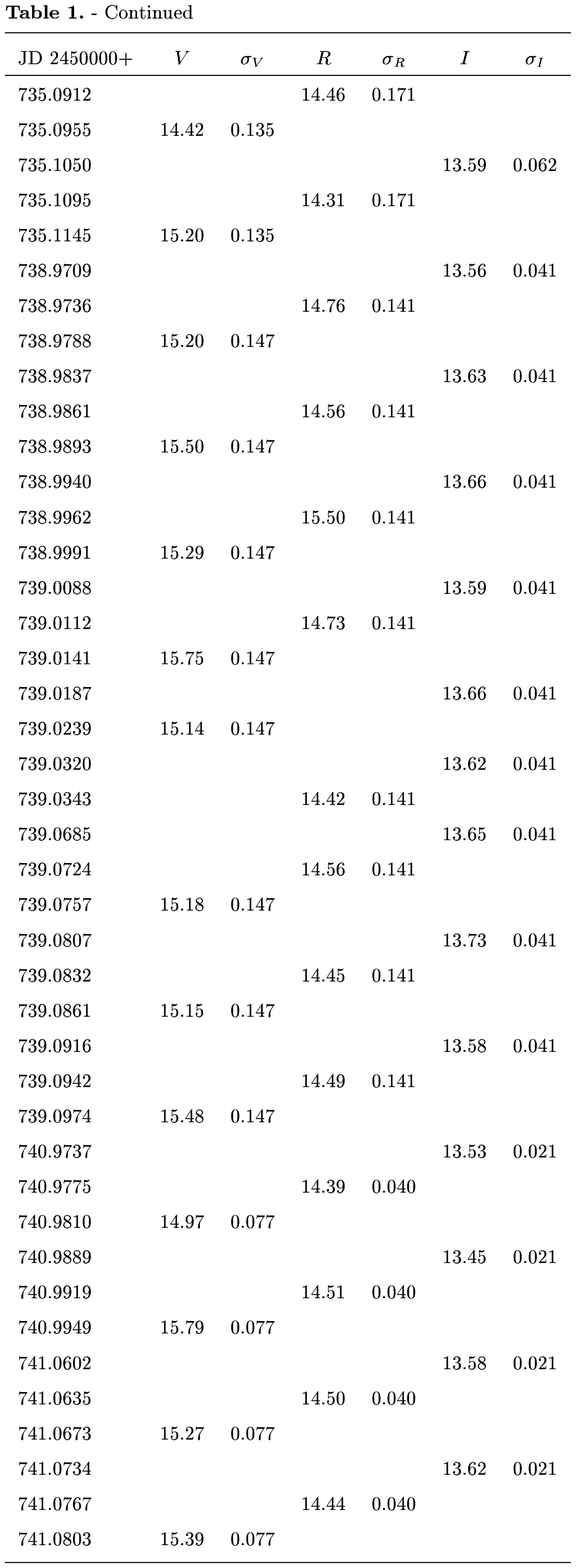}
\end{figure*}

\begin{figure*}
\vbox to7.2in{\rule{0pt}{7.2in}}
\includegraphics{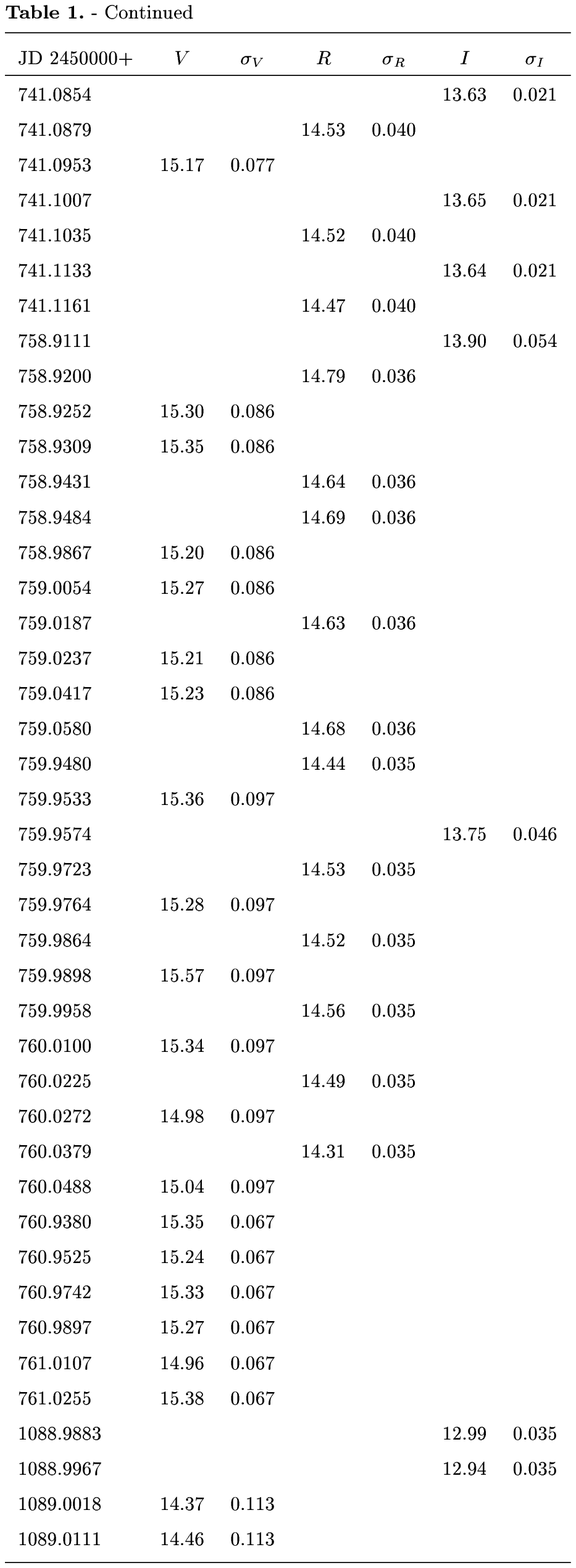}
\end{figure*}
\begin{figure*}
\vbox to7.2in{\rule{0pt}{7.2in}}
\includegraphics{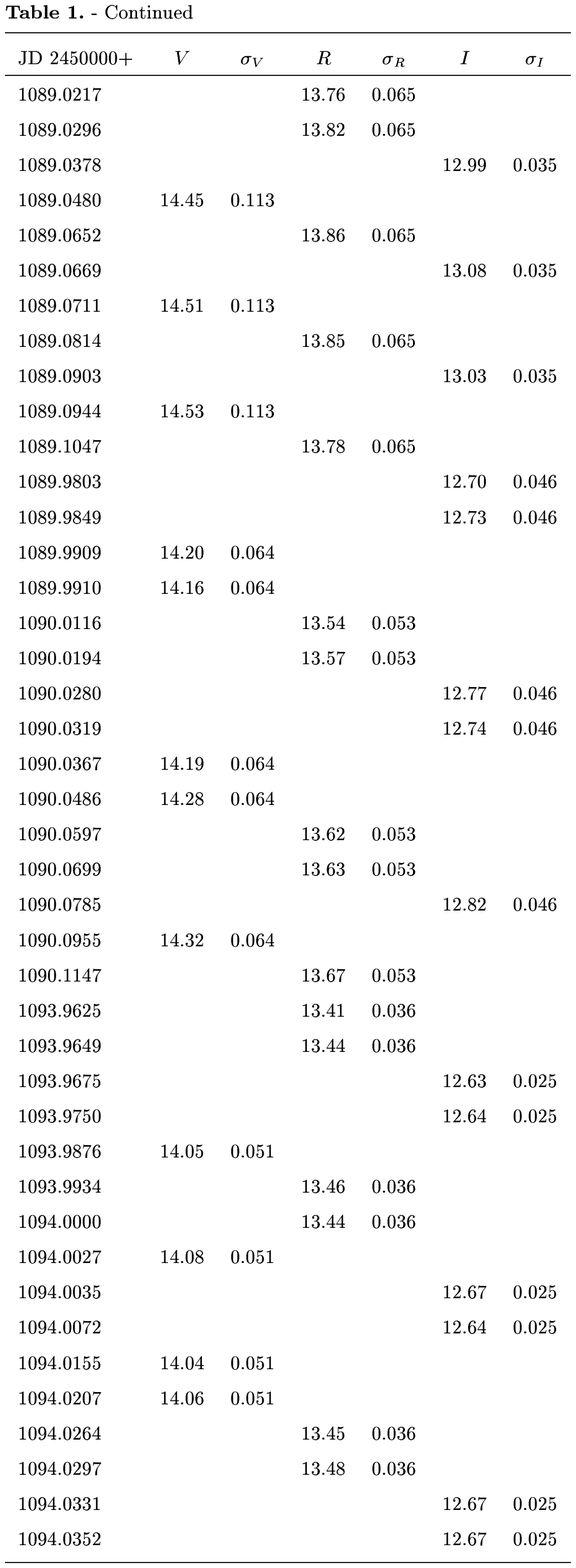}
\end{figure*}

\begin{figure*}
\vbox to7.2in{\rule{0pt}{7.2in}}
\includegraphics{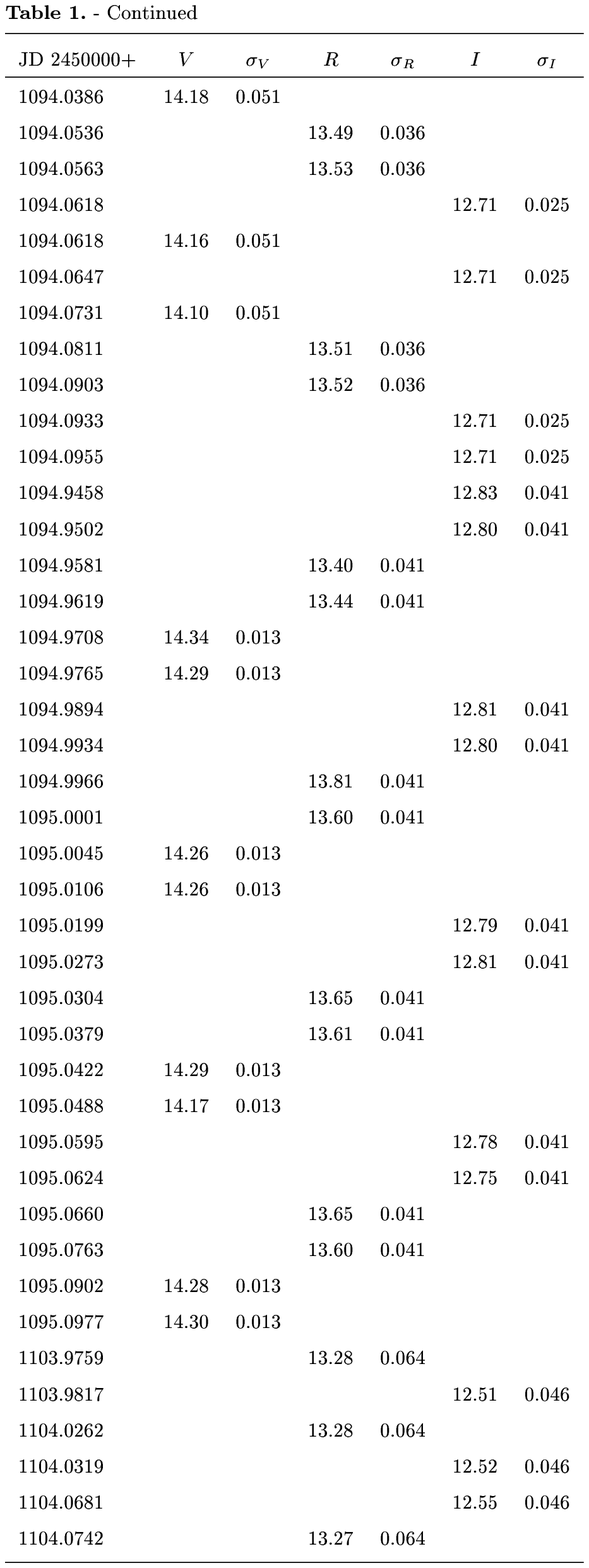}
\end{figure*}

\begin{figure*}
\vbox to7.2in{\rule{0pt}{7.2in}}
\includegraphics{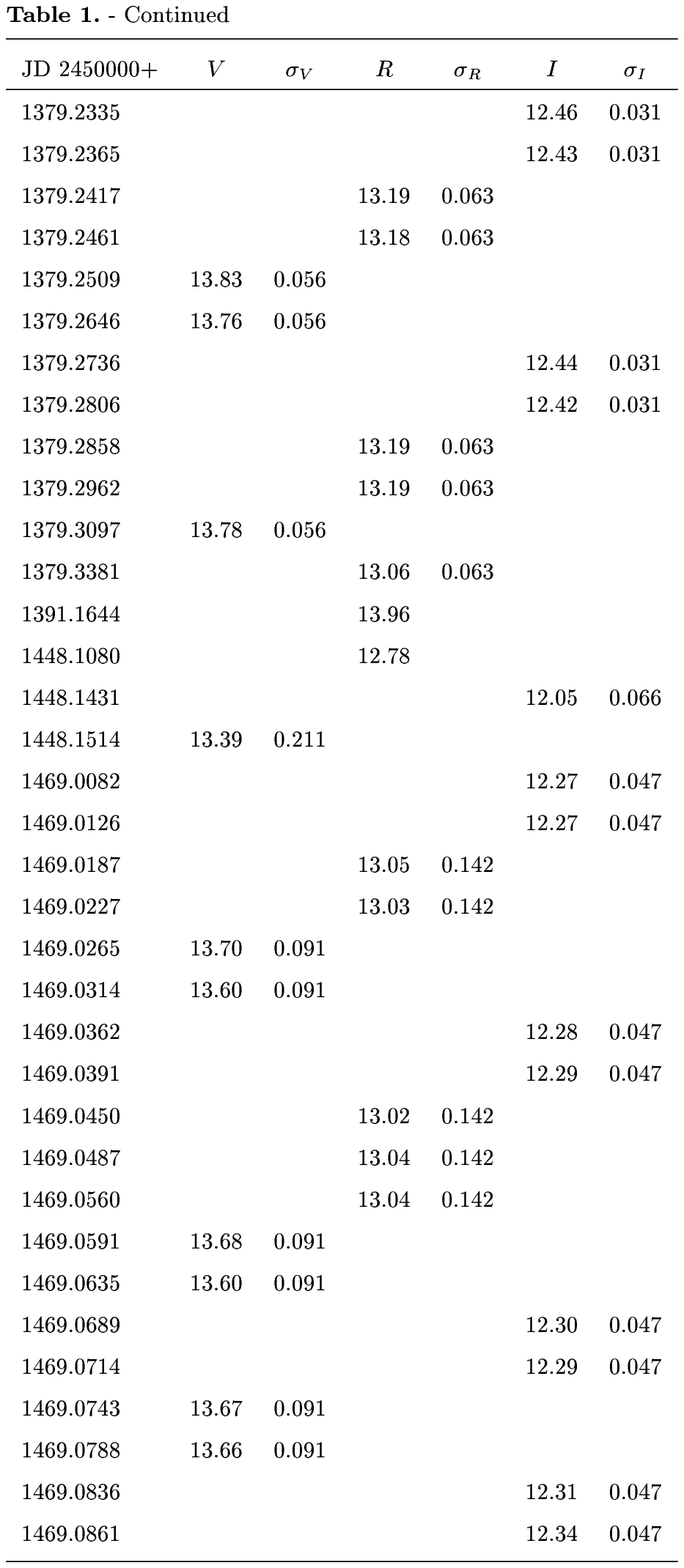}
\end{figure*}

\end{document}